# scientific reports

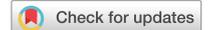

**OPEN**

# Impact of visuomotor feedback on the embodiment of virtual hands detached from the body

Sofia Seinfeld[1,2]✉ & Jörg Müller[1]

It has been shown that mere observation of body discontinuity leads to diminished body ownership. However, the impact of body discontinuity has mainly been investigated in conditions where participants observe a collocated static virtual body from a first-person perspective. This study explores the influence of body discontinuity on the sense of embodiment, when rich visuomotor correlations between a real and an artificial virtual body are established. In two experiments, we evaluated body ownership and motor performance, when participants interacted in virtual reality either using virtual hands connected or disconnected from a body. We found that even under the presence of congruent visuomotor feedback, mere observation of body discontinuity resulted in diminished embodiment. Contradictory evidence was found in relation to motor performance, where further research is needed to understand the role of visual body discontinuity in motor tasks. Preliminary findings on physiological reactions to a threat were also assessed, indicating that body visual discontinuity does not differently impact threat-related skin conductance responses. The present results are in accordance with past evidence showing that body discontinuity negatively impacts embodiment. However, further research is needed to understand the influence of visuomotor feedback and body morphological congruency on motor performance and threat-related physiological reactions.

The feelings of owning and controlling a body are critical for effective interaction with the physical environment. Recently these concepts have gained importance in immersive Virtual Reality (VR) applications, where users can experience the illusory perception of ownership and agency for a virtual body[1]. These illusions rely on brain mechanisms that build and update body representations based on real-time multisensory integration processes[2,3]. When the brain receives congruent synchronous visual, tactile, motor, or proprioceptive information with respect to a fake body or limb (i.e. mannequin, robot, or virtual arm), it resolves such sensory conflict by assuming that the artificial body is part of the real body[4,5].

In the rubber hand illusion, participants report the feeling that a fake hand is part of their body[6]. This is accomplished by placing a rubber hand in the participant's vision, while their real hand is hidden from view. The illusion is induced when the real and rubber hands are placed in an anatomically congruent position and stroked at the same time and location. After some seconds of delivering this type of sensory stimulation, participants feel body ownership of the rubber hand, evidenced by subjective reports and the occurrence of threat-related responses when somebody tries to harm the rubber hand[6,7]. Analogously, in the virtual hand illusion, participants experience feeling that a virtual hand is part of their own real body[8]. The inclusion of body tracking technologies even enables participants to control the fake virtual body through their own real-time movements[5,9]. This typically leads to the experience of embodiment in a virtual body, which is thought to be comprised of three main sub-components: (1) feeling that the virtual body is part of the real body—sense of body ownership, (2) feeling of being responsible for controlling the virtual body—sense of agency, and (3) feeling that one is located at the same position of the seen virtual body—sense of self-location[10]. Moreover, the self-attribution of an artificial body and the strength of bodily illusions have been consistently related to the processing of body-related congruent multisensory information in brain regions such as the motor cortex[11–14], the extrastriate body area[15], and intraparietal regions[16–19].

A body of evidence has shown that several factors impact the experience of body ownership or embodiment of an artificial limb[3]. For example, the strength of these perceptual illusions can be influenced by the anthropomorphic characteristics of the body part[20] and by the degree of visual realism[21]. In this regard, there is evidence that visual discontinuity of the virtual hand can negatively impact illusory body ownership. Perez-Marcos et al.[22] found that feelings of ownership decreased depending on the connectivity of the virtual hand to the rest of the virtual body. Similarly, Tieri et al.[23] and Tieri et al.[24] showed that even a small discontinuity between a virtual static hand and forearm decreased ownership, vicarious agency, and physiological responses to a threat, whereas ownership seems to be preserved when the disconnection of the arm is occluded by a black virtual rectangle.

[1]Chair of Applied Computer Science VIII, Institute of Computer Science, University of Bayreuth, 95447 Bayreuth, Germany. [2]Present address: Universitat Politècnica de Catalunya, Centre de la Imatge i la Tecnologia Multimèdia (CITM), Barcelona, Spain. ✉email: sofia.seinfeld@uni-bayreuth.de

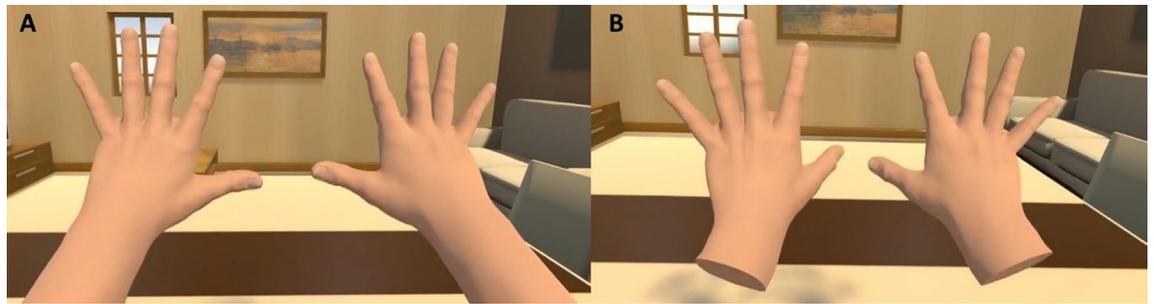

**Figure 1.** In this study, participants interacted in Virtual Reality (VR) either through Connected (**A**) or Disconnected (**B**) virtual hands.

These studies requested participants to remain still and passively observe a static[22,24] or moving collocated virtual body[23], hence they mainly exploited proprioceptive congruence to induce a body ownership illusion. However, recent studies have assessed the impact of virtual disconnected hands under the presence of visuomotor feedback, finding contradictory results. Brugada-Ramentol et al.[25] found no difference in body ownership scores between virtual hands attached to or detached from a virtual body, when the participants were able to actively control the motion of the virtual hands. Similar results were obtained in Tran et al.[26], where no differences in ownership and sense of agency were found between virtual connected or disconnected hands. However, these authors found lower performance in a selection task when controlling a virtual hand with a rendered arm, compared to doing the task with a disconnected hand. Such results stand in contrast with a study that found higher crossmodal congruency effects during the control of virtual connected hands compared to disconnected hands[27]. However, these studies did not include a more objective measure of embodiment such as the one provided by physiological response to a virtual threat[28]. Furthermore, aside from the study of Brugada-Ramentol et al.[25], the visuomotor feedback provided between the real and virtual hands was quite minimal. This means that participants were not able to control all the movements of the virtual hands, such as finger movements.

To our knowledge, no study has critically evaluated the impact of visual body disconnection for artificial hands controlled through rich visuomotor feedback (i.e., including whole hand movements as well as finger movements) on physiological responses to a virtual threat and on motor performance in an interactive bimanual task. Such evidence might prove important in order to clarify the contradictory results found by past studies, as well as to better understand the role played by top-down and bottom-up factors in the perception and construction of body representation. Moreover, several VR applications represent the user by means of virtual disconnected hands that they can move and control. For example, disconnected hands have been used for virtual bimanual assembly tasks[29], archaeological exploration[30], medical training[31], flight simulations[32], and VR keyboard inputs[33]. Since these VR applications can potentially be used in real-life training and rehabilitation contexts, it is important to understand the impact of interacting with artificial hands that are disconnected from the body on motor behavior, threat-related responses, and embodiment.

## Experiment 1

**Experimental design.** We designed a within-group experiment, including one independent variable based on the virtual *Body Visual Continuity*. The independent variable had two levels, with participants interacting in an immersive virtual environment either using Connected (Fig. 1A) or Disconnected (Fig. 1B) virtual hands (see Supplementary Video 1 for details on the visual appearance of each experimental condition). Except for the visual disconnection of the hand (i.e., the arm was not rendered), both virtual hand models were exactly the same (Fig. 1). Since each participant experienced both conditions, the *Order* of presentation was fully counterbalanced. We also matched the number of females and males assigned to each possible order.

**Participants.** A total of 31 participants (mean age = 24.10, age SD = 3.84, 22 males, 27 right-handed) took part in the study. Inclusion criteria included not suffering from sensory impairments, no neurological diseases, and no intake of psychoactive medications. The sample sizes used in Experiments 1 and 2, were determined in order to be similar to the samples sizes used in the studies of Tieri et al.[23,24]. This study was granted ethical approval by the ethical committee of the University of Bayreuth and followed ethical standards according to the Helsinki Declaration. Written informed consent was taken from all participants and they all received economic compensation for participating in the study.

**Experimental setup.** *VR scene.* The VR scene used consisted of a virtual room with a table, television, sofa, window, and a couple of paintings. It was programmed using Unity 3D and experienced through an HTC Vive head-mounted display (HMD). Hand and finger tracking was enabled using a Leap Motion sensor, which provided real-time visuomotor feedback matching the participants' real hand movements with those made by the virtual hands.

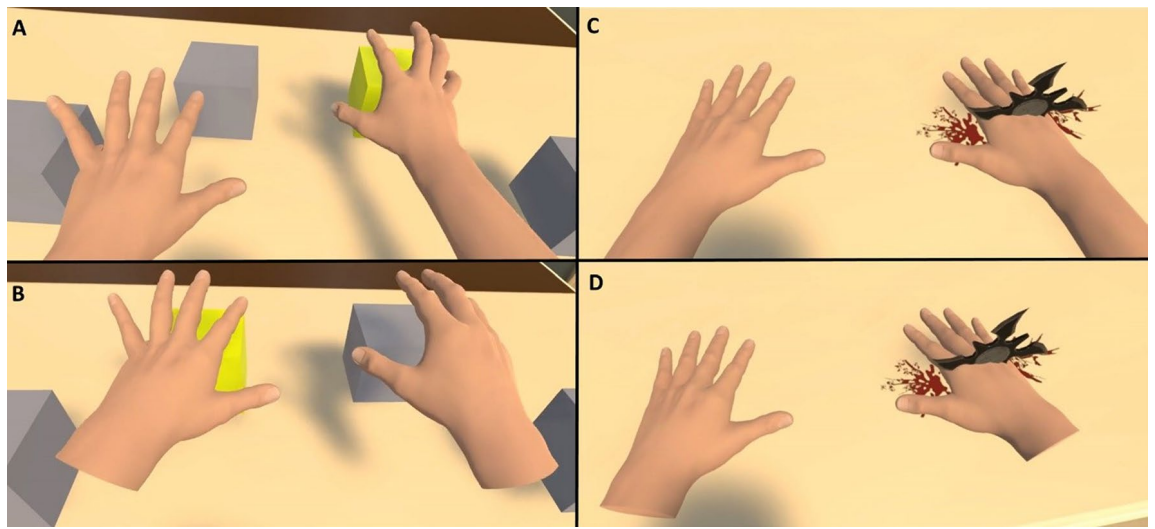

**Figure 2.** Participants executed the virtual motor task, based on touching yellow-colored cubes, through Connected (**A**) and Disconnected (**B**) virtual hands. At the end of the study, the virtual Connected (**C**) and Disconnected (**D**) were threatened with a virtual shuriken.

The Connected hand models consisted of virtual hands including a forearm, connected to a full virtual body (Fig. 1A). The full virtual body position was aligned with respect to the participant's real physical body, based on head and hand position. Therefore, there was no perceptible proprioceptive mismatch between the virtual and real body. In the Disconnected condition, we used exactly the same hand models used for the Connected hand condition, but in this case the forearm and rest of the virtual body were invisible (Fig. 1B). Thus, participants just saw a pair of virtual hands floating in mid-air which were collocated with their own real hands. They were also able to control the virtual hands in real-time based on their own hand movements. The specific visual appearance of the virtual hands was designed based on the results of pilot studies where we validated that the visual discontinuity (i.e., lack of virtual forearm) or continuity of the hands was clearly visible throughout the experiment. Hence, while exploring the virtual environment, performing the motor task, or looking at the virtual threat, participants could clearly see the virtual connection or disconnection of the hands as it can be seen in Fig. 1 and the Supplementary Video 1.

*Motor task.* We designed a task based on the quick selection of targets (i.e., virtual cubes that lit up in yellow) by touching them, in order to assess possible differences in motor performance based on the connection or disconnection of the virtual hands. During the virtual experience, participants saw four grey cubes placed within arm's reach on a table in front of them (Fig. 2A,D). In a random order, one of the cubes turned yellow (i.e., the target) and participants were instructed to touch the highlighted cube as quickly as possible. As soon as the target was touched, its color changed back to grey and another cube in a different location turned yellow instead. Participants were explicitly instructed to carry out the task using both hands. The experimenter could see in real time the participant's view of the virtual environment through a screen. Through these real-time observations of the VR scene, it was confirmed that participants used both of their hands to perform the task and it was also verified that most participants used their right hand to touch targets located at the right side and their left hand to touch targets located at the left side of the virtual table. On a technical level, touching of targets was triggered by the built-in collider system of Unity. A BoxCollider was placed in the virtual hands. When one of the colliders intersected with a target's BoxCollider, a touch event was triggered. The task was configured with the constraint that the same target could not light up twice in a row, thus each time a target was touched the next cube that would light up was in a different location. In each experimental condition, the participants had to perform this task for a total duration of 90 s. See the Supplementary Video 1 for more details on the motor task used in the virtual reality setup.

*Virtual threat.* For the threat scene, we used a virtual shuriken object that fell from above onto the artificial virtual hands. When the threat reached the hand, it stopped and virtual blood was rendered (Fig. 2C,D). Audio feedback of the falling shuriken was included.

**Measures.** *VR questionnaire.* We included a series of questions addressing different aspects related to the VR experience (all questions can be seen in Table 1). The questions were answered on a 7-point Likert scale, where 1 meant "completely disagree" and 7 "completely agree" with the statement. More specifically, the questions *Ownership*, *NotMine*, and *TwoHands* were related to the experience of a body ownership illusion and inspired by the original rubber and virtual hands illusion studies[6,8]. The *Agency* and *Control* questions were meant to assess subjective perceptions of feeling responsible for moving and controlling the virtual hands based on the provided visuomotor correlations between the virtual and the real hands. We also included a specific question to measure whether participants felt that their own body was located where they saw the virtual body, namely *SelfLoc*. Finally, the questions *Threat* and *Real* were related to the overall degree of immersion experienced in the VR scene. All included questions were selected from previous similar studies, as explained in the study by Gonzalez-Franco et al.[34]. Participants completed the questionnaire immediately after each experimental condition.

| Variable | Questionnaire item |
|---|---|
| *Ownership* | I felt that the virtual hands were my own hands |
| *NotMine* | I felt as if the virtual hands I saw were someone else's |
| *TwoHands* | It seemed as if I might have more than two hands |
| *Agency* | I felt like I could control the virtual hands as if they were my own hands |
| *Control* | I felt as if the movements of the virtual hands were caused by my movements |
| *SelfLoc* | I felt as if my hands were located where I saw the virtual hands |
| *Threat* | I felt threatened by the shuriken (knife attacking my hand) although I knew it was virtual |
| *Real* | I felt the experience was real, although I knew it was virtual |

**Table 1.** Questionnaire items included in the VR questionnaire.

*Skin conductance response (SCR) to a virtual threat.* Several studies have established the relationship between body ownership and the reflex of trying to protect the artificial limb from a threat. For instance, feelings of body ownership for a virtual hand that is threatened leads to an enhancement of motor cortex activations[28], skin conductance[24], and heart rate deceleration[9]. Based on these reasons we gathered preliminary data on threat-related skin conductance responses to a virtual shuriken which harmed the virtual Connected (Fig. 2C) and Disconnected hand (Fig. 2E), respectively. We only included one single virtual threat trial (i.e., shuriken harming the virtual hand) per experimental condition, thus rendering our results as preliminary. The decision of recording a single threat trial per condition was done with the aim of controlling for possible habituation effects due to the repeated exposure to the same threatening stimuli[35]. Since the inclusion of more trials is needed to obtain more conclusive evidence and to observe whether results are replicated in different repetitions of the virtual harm, all details related to SCR (i.e., recording, analysis, and results) are presented in the Supplementary Information 2 and should be interpreted with caution.

*Motor performance.* We evaluated motor performance based on a quick bimanual task, which required participants to quickly select targets (see details in the *Motor task* section). Motor performance was computed as the number of Hits (i.e., touched targets) in 90 s. This was calculated for each experimental condition, respectively. The motor performance data of Experiments 1 and 2 was extracted using MATLAB.

*Final interview.* Upon the completion of the study, we carried out a short informal interview, to better understand the participants' perceptions. No explicit reference to the connection or disconnection of the virtual hands was made in the interview or throughout the experiment, in order not to bias the participants' responses. This interview was comprised of two main open questions, which were: *Question 1*) What are your overall feelings and thoughts about the VR scenes you just experienced? And *Question 2*) Did you notice any difference between the first and second time you went through the VR experience? If yes, describe what differences you noticed? The responses in this interview were used subsequently to define the factor *Awareness* which was included as a factor in several of our statistical analyses. Through the interview answers we noticed that while some participants reported *Being Aware* of the experimental manipulation (i.e., differences in the hand appearance between the Connected and Disconnected hands), other participants reported *Not Being Aware* of any difference between the experimental conditions.

**Procedure.** Participants were given information about the study and signed a consent form if they were willing to participate. They were randomly assigned to one of the possible experimental orders of the conditions. Before the study started, we placed the physiological recording equipment on the participants.

Participants were requested to sit down in front of a table. When the VR scene started, they saw themselves immersed in a virtual living room that also had a virtual table located in front of them. They also saw a virtual counterpart of their hands, which moved accordingly to their real hand movements. In one experimental condition, they saw virtual hands that were Connected to their virtual body and included arms (Fig. 1A). In the other condition, participants only saw a pair of virtual collocated Disconnected hands not including arms (Fig. 1B; details given in *VR Scene* section). First, we familiarized participants with the virtual scene by asking them to describe their surroundings (i.e., the virtual room) and their new virtual bodies. Here we explicitly asked participants to look down towards their virtual body and look at their hands. Subsequently, we asked participants to relax, remain silent, and breathe slowly for 2 min, to record a baseline measure of their physiological state. When the recording of the physiological baseline ended, the scene vanished (i.e., turned black).

Here we explained to participants that when the scene appeared again, they would have to perform the virtual motor task (see detailed description in the section *Motor task*). We instructed them to touch the targets (i.e., yellow virtual cubes) as accurately and as fast as possible using both hands, until the scene automatically stopped.

When participants finished the virtual motor task, the scene vanished and turned black. Here, without visual inputs, the experimenter placed the participants' hands on top of the real table. When the scene reappeared, the participants saw their corresponding virtual hands (Connected or Disconnected) resting on a virtual table. The virtual hands' position corresponded with the users' real hands' position on the table. We instructed participants to observe their right virtual hand. After 10 s, the virtual threat was triggered, and a shuriken fell from the top harming the virtual hand. Participants remained looking at the hand for further 30 s while we kept recording their physiological reaction. The experimenter ensured that the participants were looking at their right virtual hands through a screen where it was possible to visualize the participants' real-time view of the virtual environment. Then we asked participants to take off the HMD and to complete the VR questionnaire basing their responses on the overall virtual scenes they just experienced, including the familiarization phase, their perceptions during the motor task, and the virtual threat part.

After the completion of the first experimental condition, the same procedure was repeated for the second experimental condition. The only difference between the conditions was the virtual hand model used.

**Statistical analysis approach.** The VR questionnaire items were analyzed using multilevel mixed-effects ordered logistic regressions using Stata 16.1 software. The questionnaire items data met the assumptions for ordinal logistic regressions analyses since the dependent variables are ordinal (i.e., questionnaire), the independent variables are categorical (i.e., *Body Visual Continuity* and *Awareness*), there is no multi-collinearity (i.e., automatically checked by Stata), and Brant tests carried out in each repeated measures questionnaire items showed that the assumption of proportional odds is not violated.

In the multi-level ordered logistic regressions, *Body Visual Continuity* (i.e., Connected or Disconnected hands), *Awareness* (i.e., Being Aware or Not Aware about the experimental manipulation) and their interaction term were introduced as fixed factors. Participants' IDs were set as random effects in the model in order to control for the within-group nature of the experiment. We decided to include the additional factor *Awareness* in the analysis to control for the potential influence of participants consciously noticing the visual connection/disconnection of the hand during the experiment or not noticing (see details of the *Awareness* factor classification in "Final Interview"). A total of 16 participants clearly noticed the experimental manipulation during the experiment (i.e., visual connection/disconnection of the hand), while 15 participants did not consciously notice this difference during the study.

Motor performance data was analyzed using mixed-design ANOVAs carried out using SPSS Version 24, where the factor *Body Visual Continuity* was included as a within-group factor and the factors *Awareness* and *Order* were set as the between-group factors. The factor *Awareness* was included in the analysis to control for the potential influence of participants consciously noticing the experimental manipulation (i.e., visual continuity of the body) or not. Moreover, despite the use of a full counterbalanced design, we further included the factor *Order* in this analysis to thoroughly control for physiological habituation effects related to the repeated exposure to the virtual threat[35] and to control for potential learning effects (i.e., training) in the motor task[36]. The residual errors of the ANOVA analysis were tested for normality using Shapiro–Wilk tests. Significance of results was calculated with a 95% confidence level. Post-hoc comparisons for significant interaction effects were based on paired comparison t-tests including Benjamini and Hochberg corrections for multiple comparisons[37]. Figures of the data were plotted using Stata 16.1.

**Results.** *VR questionnaire.* A significant effect of the factor *Body Visual Continuity* was found in *Ownership* (Coef = − 1.39, z = − 1.99, p = 0.046, 95% CI − 2.75 to − 0.20), showing that participants reported higher body ownership scores in the Connected (Median (Mdn) = 5, Interquartile Range (IQR) = 1) condition compared to the Disconnected one (Mdn = 4, IQR = 2). Similarly, participants also reported a higher feeling of *Control* (Coef = − 1.89, z = − 2.44, p = 0.02, 95% CI − 3.40 to -0.37) over the virtual hands when interacting through Connected hands (Mdn = 6, IQR = 1) compared to Disconnected hands (Mdn = 6, IQR = 1). Finally, the sense of *SelfLoc* (Coef = − 2.18, z = − 2.27, p = 0.02, 95% CI − 4.06 to − 0.30) was also stronger for Connected hands (Mdn = 6, IQR = 1) compared to Disconnected hands (Mdn = 6, IQR = 2). In the *NotMine* questionnaire item we found a significant main effect of *Body Visual Continuity* (Coef = 1.87, z = 2.50, p = 0.01, 95% CI 0.41–3.33), indicating that scores were lower for the Connected hands (Mdn = 2, IQR = 2) compared to the Disconnected hands (Mdn = 3, IQR = 2). No significant interaction effects between *Body Visual Continuity*\**Awareness* or main effects of *Awareness* were found in any of the questions. Finally, no significant main effects of *Body Visual Continuity* were found in the questions *Agency*, *TwoHands*, *Threat*, and *Real,* showing that the Connected and Disconnected hands did not differ in these aspects. Detailed results of the ordered logistic regressions are given in Table 2. Figure 3 shows boxplots of all questionnaire items.

*Motor performance.* In relation to the number of cubes touched in 90 s, we found a significant main effect of *Body Visual Continuity* ($F(1,26) = 5.03$, p = 0.03, partial $\eta^2 = 0.16$) and an interaction effect between *Body Visual Continuity*\**Awareness* ($F(1,26) = 8.87$, p < 0.01, partial $\eta^2 = 0.25$). This interaction effect indicated that only in the case where participants did not explicitly notice the experimental manipulation (i.e., visual continuity of the hand), they touched a significantly higher number of targets in 90 s with the Connected hands [mean (M) = 167.93, standard deviation (SD) = 22.29] compared to the Disconnected hands (M = 142.13, SD = 29.30) (t = 3.90, df = 14, p < 0.01). However, no significant difference between Connected (M = 141.13, SD = 35.10) and Disconnected M = 150.47, SD = 37.48) hands was found in participants who were aware of the experimental manipulation during the study (t = −1.27, df = 14, p = 0.23), and thus who did not notice the difference between the hand models (see Fig. 4). No significant main effects of *Awareness* (F(1,26) = 1.05, p = 0.32, partial $\eta^2$ = 0.04) and *Order* (F(1,26) = 0.32, p = 0.58, partial $\eta^2$ = 0.01) were found. Moreover, the interaction between *Body Visual Continuity*Order* (F(1,26) = 0.19, p = 0.66, partial $\eta^2$ < 0.01) was not significant. The residual errors of the ANOVA were normally distributed.

| Item | Body visual continuity | | | | | | Awareness | | | | | | Interaction BVC*A | | | | | |
|---|---|---|---|---|---|---|---|---|---|---|---|---|---|---|---|---|---|---|
| | Coef | SE | z | p | CI | OR | Coef | SE | z | p | CI | OR | Coef | SE | z | p | CI | OR |
| Ownership | −1.39 | 0.70 | −1.99 | 0.04 | −2.75 / −0.02 | 0.25 | −0.22 | 0.76 | −0.28 | 0.78 | −1.77 / 1.27 | 0.81 | 0.88 | 0.95 | 0.92 | 0.36 | −0.99 / 2.74 | 2.40 |
| NotMine | 1.87 | 0.75 | 2.50 | 0.01 | 0.41 / 3.33 | 6.49 | 0.17 | 0.93 | 0.18 | 0.86 | −1.65 / 1.98 | 1.18 | −1.31 | 1.00 | −1.30 | 0.19 | −3.26 / 0.66 | 0.27 |
| TwoHands | 1.22 | 1.03 | 1.18 | 0.24 | −0.80 / 3.24 | 3.38 | 0.63 | 1.59 | 0.40 | 0.69 | −2.48 / 3.74 | 1.88 | −1.60 | 1.50 | −1.06 | 0.29 | −4.55 / 1.35 | 0.20 |
| Control | −1.89 | 0.77 | −2.44 | 0.02 | −3.40 / −0.37 | 1.51 | 0.16 | 0.89 | 0.18 | 0.86 | −1.59 / 1.90 | 1.17 | 1.29 | 1.03 | 1.25 | 0.21 | −0.74 / 3.32 | 3.63 |
| Agency | −0.48 | 0.80 | −0.59 | 0.55 | −2.05 / 1.10 | 0.62 | −0.22 | 1.00 | −0.22 | 0.82 | −2.18 / 1.74 | 0.80 | 1.18 | 1.17 | 1.01 | 0.31 | −1.11 / 3.48 | 3.27 |
| SelfLoc | −2.18 | 0.96 | −2.27 | 0.02 | −4.06 / −0.30 | 0.11 | −0.54 | 1.78 | −0.31 | 0.76 | −4.03 / 2.94 | 0.58 | −0.26 | 1.21 | −0.21 | 0.83 | −2.63 / 2.12 | 0.77 |
| Threat | −0.56 | 0.68 | −0.82 | 0.41 | −1.89 / 0.78 | 0.57 | 0.28 | 1.14 | 0.25 | 0.80 | −1.95 / 2.52 | 1.33 | −0.04 | 0.98 | −0.04 | 0.97 | −1.96 / 1.88 | 0.96 |
| Real | −0.90 | 0.67 | −1.34 | 0.18 | −2.22 / 0.42 | 0.41 | 0.26 | 1.05 | 0.24 | 0.81 | −1.81 / 2.32 | 1.29 | 0.53 | 0.97 | 0.55 | 0.58 | −1.37 / 2.24 | 1.71 |

**Table 2.** Experiment 1 multilevel ordered logistic regressions analyses of the VR questionnaire items. In the analyses, Body Visual Continuity (BVC), Awareness (A), and their interaction were set as fixed factors and participants´ IDs defined as random effects of the model. The table shows the Coefficient (Coef), Standard Error (SE), z-values, p-values 95% Confidence Intervals (CI), and Odds Ratios (OR) for the ordered logistic regressions performed for each questionnaire item.

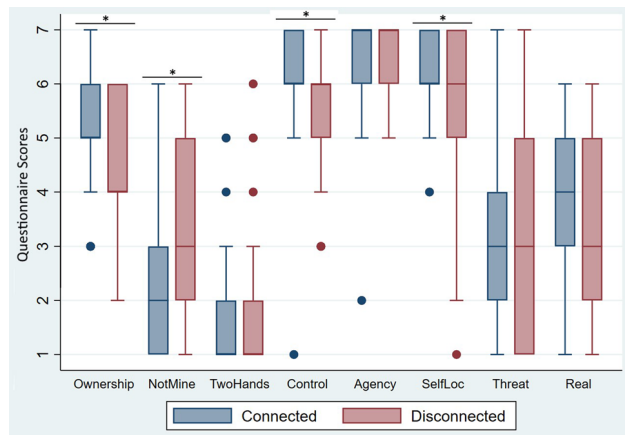

**Figure 3.** Experiment 1 boxplots for the VR questionnaire showing the median, interquartile ranges, maximum scores, and minimum scores for each questionnaire item in the Connected and Disconnected hands experimental conditions, respectively. Dots represent outlier values and the asterisks show the questionnaire items in which a significant difference between the Connected and Disconnected hands condition was found.

*Final interview.* In response to *Question 1* (i.e., What are your overall feelings and thoughts about the VR scenes you just experienced?), several participants reported that they had fun and felt that the touching targets task was engaging. Moreover, other participants also described that they were surprised by the virtual threat and that although they knew it was virtual, at that moment they felt the reflex of protecting their virtual hand.

Based on participants' answers to *Question 2* (i.e., Did you notice any difference between the first and second time you went through the VR experience?), we found that only 16 participants explicitly mentioned the inclusion (Connected) or exclusion (Disconnected) of the virtual forearm as a clear difference between the two VR experiences. In contrast, 15 participants did not mention the Connection or Disconnection of the virtual hands in response to *Question 2*. This means that almost half of the participants were not consciously aware of the experimental manipulation (i.e., differences in virtual hand appearances). In order to control for participants being (n = 16) or not being aware (n = 15) of the experimental manipulation, the variable *Awareness* was added as a factor to control for in all the statistical analyses carried out.

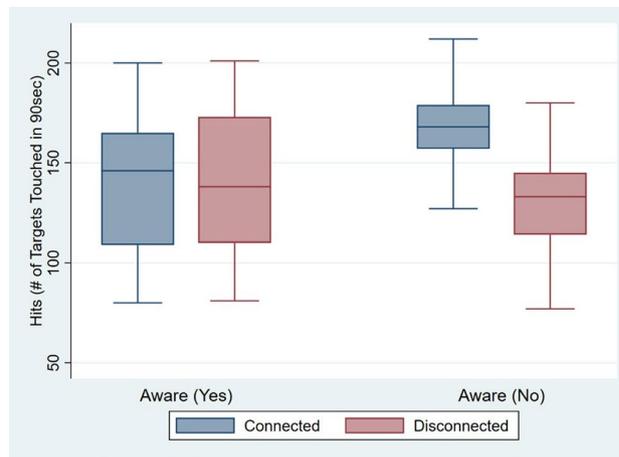

**Figure 4.** Boxplots of the number of hits (i.e., targets touched in 90 s) achieved in the Connected and Disconnected hands conditions when participants were aware or not of the experimental manipulation in Experiment 1. The boxplots show the median, interquartile ranges, maximum scores, and minimum scores for the number of hits.

It should be noted that some of the participants who did not mention the connection or disconnection of the hands did unknowingly indicate that they preferred the session with the Connected hand condition. As an example of one of these cases, one participant answered: "*I liked the first VR session more, because it looked more realistic*". In this case, the first session corresponded to the Connected hand condition. However, when asked explicitly whether he noticed any difference between the sessions, the same participant answered that he could not actually tell what the specific differences between the VR scenes were. Several participants had similar responses.

Almost all participants who explicitly noticed the differences in the visual appearance of the hands reported that they felt better when interacting with the virtual Connected hands. Here we list some examples of the most illustrative phrases expressed by the participants:

Participant 1: "*The unconnected hands appeared less realistic, less accurate, and less comfortable.*"
Participant 2: "*The connected hands felt more realistic and I felt as if I was slower with the disconnected hands.*"
Participant 3: "*The hands with arms were more realistic and the movements were more accurate.*"

**Discussion.** We found that on a subjective perceptual level, participants reported stronger feelings of body ownership, control, and self-location towards a virtual body including fully rendered connected hands compared to disconnected hands. However, the disconnection of the hands did not diminish the sense of agency or feelings of being responsible for the movements made by the virtual hands. Moreover, in this study we found results indicating that participants' motor performance was enhanced in the connected hand condition compared to the disconnected condition, especially in those cases where participant did not report being consciously aware of the difference in the visual appearance of the virtual hands (i.e., connected or disconnected). However, since we did not record the position of the targets being touched and since the order in which the targets lit up was randomly generated, we cannot disregard that the differences between conditions in motor performance might be explained by the degree of task difficulty. In other words, it is possible that in some conditions targets that were located closer to the participant (front) lit up more frequently that targets located further away (back; Fig. 5). In order to control for these possible confounding variables in the motor task, we carried out Experiment 2. With this new experiment we also aimed to see whether we could replicate the results found in the self-reported VR questionnaire. Finally, although it should be interpreted with caution since these results are only based on one threat trial per condition, our preliminary data on SCR to a virtual threat indicated that participants had physiological responses related to anxiety (i.e., increase in skin conductance) when the virtual hands were threatened, independently of the observation of body visual discontinuity (see Supplementary Information 2).

## Experiment 2
**Introduction.** In Experiment 2 we further researched the impact of visual body discontinuity on embodiment and motor performance. Specifically, Experiment 2 was very similar to Experiment 1, with the difference that in this new study we recorded the location of the touched targets (i.e., cubes that lit up) and ensured that the level of difficulty of the motor task was equivalent between the experimental conditions.

**Experimental design.** Experiment 2 followed the same experimental design as the one described in Experiment 1. A fully counterbalanced within-groups design experiment was carried out. The independent variable of this study was *Body Visual Continuity*, with participants interacting in an immersive virtual environment either using Connected or Disconnected virtual hands.

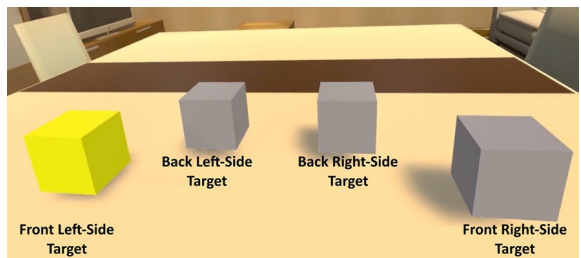

**Figure 5.** In Experiment 1 and 2, participants had to quickly touch a target that lit up from four different possible cubes. Two cubes were placed to the right side of the participants and two to the left side. Moreover, two cubes were closer to the participant's virtual body (front) and two further away (back). In this specific figure, the target that participants are required to quickly touch is represented by the front left-side cube which is lit up in yellow.

**Participants.** A total of 20 participants (mean age = 26, age SD 4.03, 12 males, 19 right-handed) took part in the study. Inclusion criteria were the same as described in Experiment 1. This study was granted ethical approval by the ethics committee of the University of Bayreuth and followed ethical standards according to the Helsinki Declaration. Participants received economic compensation for their participation. Written informed consent was taken from all the participants.

**VR scene and setup.** The virtual environment used in Experiment 2 was very similar to that of Experiment 1. Participants were immersed in a virtual room including a table, television, sofa, window, and a couple of paintings. In this scene, participants also experienced a virtual body from a first-person perspective, either including hands that were connected to the rest of the virtual body (Fig. 1A) or detached from the body (Fig. 1B). During the experiment participants executed the same motor task as in Experiment 1, consisting in the quick selection of targets with both hands (Fig. 2A,B). However, in this study we carefully controlled for the position of the touched targets (Fig. 5) and ensured that the level of task difficulty was identical across conditions, aspects that we did not control for in Experiment 1. This was achieved by: (1) making the cubes light up in the same order in the Connected and Disconnected conditions, (2) assigning the same probability to the four cubes of becoming a target, and by (3) providing a training phase. Therefore, in this experiment each of the four cubes could light up a total of 40 times, resulting in a total of 160 Hits (i.e. trials) per condition. For each participant we previously generated and recorded a unique pre-defined sequence that defined the order in which the cubes would light up, with the only constraint that the same cube could not light up twice in a row. Hence, using a pre-defined sequence of targets for each participant we ensured that the level of difficulty of the motor task was equivalent in the Connected and Disconnected hands, since cubes lit up following the same order in both conditions. To ensure that the results were not influenced by the use of a unique sequence, we generated different sequences (i.e., orders) for each individual participant. Moreover, to avoid the potential influence of learning effects, in this experiment we included a short training phase before the start of the actual experimental trials. The training phase also consisted in the quick selection of targets, however in this case each target lit up a total of 12 times, resulting in a total of 48 training trials per condition. The sequence in which the cubes lit up in the training phase was equivalent for all participants.

Importantly, in this new study, we calculated the time taken to touch each target from stimulus onset. Moreover, we also recorded the order in which the different targets lit up and their positions since this data was not registered in Experiment 1. As in Experiment 1, the experimenter verified that participants used both of their hands to perform the task. These observations also indicated that most participants used their right hand to touch right targets and their left hand to touch left targets.

No virtual threat was included in Experiment 2, since the main goal of this study was to assess motor performance when interacting with Connected and Disconnected hands.

**Measures.** *VR questionnaire.* In this study we also administered the same VR questionnaire shown in Table 1, immediately after participants experienced each of the experimental conditions. The only question that was not included in this study was the threat question, since in the present experiment we did not include a virtual threat.

*Motor performance.* In this experiment motor performance was computed based on the averaged reaction times taken to touch each target from stimulus onset. Since in this study we also recorded the position of the touched targets, we were also able to calculate the average time taken to touch each target depending on its vertical (left and right) or horizontal position (front or back). In Experiments 1 and 2, there was a total of four possible positions of the targets which can be seen in Fig. 5.

| Item | Body visual continuity | | | | | | Awareness | | | | | | Interaction BVC*A | | | | | |
|---|---|---|---|---|---|---|---|---|---|---|---|---|---|---|---|---|---|---|
| | Coef | SE | z | p | CI | OR | Coef | SE | z | p | CI | OR | Coef | SE | z | p | CI | OR |
| Ownership | − 2.02 | 0.99 | − 2.02 | 0.04 | − 3.97 / − 0.06 | 0.13 | 2.29 | 1.15 | 1.99 | 0.05 | − 1.77 / 3.28 | 9.84 | 0.76 | 1.29 | 0.59 | 0.56 | − 1.77 / 3.28 | 2.13 |
| NotMine | 1.34 | 1.00 | 1.34 | 0.18 | − 0.63 / 3.31 | 3.82 | − 0.42 | 1.47 | − 0.28 | 0.78 | − 3.31 / 2.47 | 0.66 | − 2.00 | 1.41 | − 1.14 | 0.16 | − 4.76 / 0.77 | 0.14 |
| TwoHands | 1.59 | 1.24 | 1.28 | 0.20 | − 0.84 / 4.03 | 4.93 | − 2.56 | 2.22 | − 1.15 | 0.25 | − 6.91 / 1.79 | 0.08 | 0.06 | 1.98 | 0.03 | 0.98 | − 3.81 / 3.93 | 1.06 |
| Control | − 3.69 | 1.25 | − 2.94 | < 0.01 | − 6.14 / − 1.23 | 0.03 | − 0.91 | 1.06 | − 0.86 | 0.39 | − 2.98 / 1.16 | 0.40 | 2.49 | 1.38 | 1.80 | 0.07 | − 0.22 / 5.19 | 12.1 |
| Agency | − 2.23 | 1.18 | − 1.90 | 0.06 | − 4.54 / 0.70 | 0.11 | 0.14 | 1.38 | 0.10 | 0.92 | − 2.56 / 2.84 | 1.15 | 1.30 | 1.48 | 0.88 | 0.38 | − 1.61 / 4.21 | 3.67 |
| SelfLoc | − 2.03 | 0.97 | − 2.08 | 0.04 | − 3.94 / − 0.12 | 0.13 | − 0.06 | 0.87 | − 0.06 | 0.95 | − 1.77 / 1.66 | 0.94 | 2.05 | 1.29 | 1.59 | 0.11 | − 0.48 / 4.57 | 7.75 |
| Real | − 2.04 | 0.96 | − 2.12 | 0.03 | − 3.93 / − 0.15 | 0.13 | − 0.03 | 1.22 | 0.03 | 0.98 | − 1.90 / 1.89 | 0.97 | 1.25 | 1.22 | 1.02 | 0.31 | − 1.15 / 3.65 | 3.48 |

**Table 3.** Experiment 2 multilevel ordered logistic regressions analyses of the VR questionnaire items. In the analyses, Body Visual Continuity (BVC), Awareness (A), and their interaction were set as fixed factors and participants´ IDs defined as random effects of the model. The table shows the Coefficient (Coef), Standard Error (SE), z-values, p-values, 95% Confidence Intervals (CI), and Odds Ratios (OR) for the ordered logistic regressions run for each questionnaire item.

*Final interview.* Upon the completion of the study, we also performed the final interview described in Experiment 1.

**Statistical analysis approach.** VR questionnaire items were analyzed using multilevel mixed-effects ordered logistic regressions. In this analysis we included as fixed factors *Body Visual Continuity* (i.e., Connected or Disconnected hands), *Awareness* (i.e., Being Aware or Not Aware of the experimental manipulation during the study), and their interaction, with random effects over participants' IDs, to take into account the within-groups nature of the experiment.

Motor performance was analyzed with a mixed-model ANOVA where the factors *Body Visual Continuity* (i.e., Connected or Disconnected), *Vertical Position of Target*s (i.e., front or back targets), and *Horizontal Position of Target* (i.e., right or left targets), were included as within-group factors. Moreover, the factors *Awareness* and *Order* were set as the between-group factors. The residual errors of the ANOVA analysis were tested for normality using Shapiro–Wilk tests. Significance of results was calculated with a 95% confidence level.

**Results.** *VR questionnaire.* A significant effect of the factor *Body Visual Continuity* was found in *Ownership* (Coef = − 2.02, z = − 2.02, p = 0.04, 95% CI − 3.97 to 0.06), indicating that participants experienced higher body ownership when interacting with Connected (Mdn = 6, IQR = 1.75) compared to Disconnected (Mdn = 5.5, IQR = 1.75) hands. Moreover, we also found that participants perceived a higher level of *Control* (Coef = − 3.69, z = − 2.94, p < 0.01, 95% CI − 6.14 to − 1.23) over the Connected hands (Mdn = 6, IQR = 1) compared to the Disconnected hands (Mdn = 5.5, IQR = 1.75). The sense of self-location (*SelfLoc*; Coef = − 2.03, z = − 2.08, p = 0.04, 95% CI − 3.94 to − 0.12) was also significantly stronger for the Connected (Mdn = 6, IQR = 1) than Disconnected hands (Mdn = 6, IQR = 1.75). The VR scene was perceived as significantly more *Real* (Coef = -2.04, z = − 2.12, p = 0.03, 95% CI − 3.93 to − 0.15) when interacting with Connected (Mdn = 6, IQR = 1) virtual hands compared to Disconnected (Mdn = 5, IQR = 2) ones. No significant main effect of *Awareness* was found in any of these questions except for *Ownership*. We found that *Ownership* scores in this case were higher in those participants who did not notice the experimental manipulation during the study (Mdn = 6, IQR = 1), compared to those who did notice (Mdn = 5, IQR = 2; Coef = 2.29, z = 1.99, p = 0.05, 95% CI − 1.77 to 3.28). However, the interaction between *Body Visual Continuity* and *Awareness* was not significant for any of the questionnaire items, demonstrating that *Body Visual Continuity* had an impact on the different questionnaire items independently from *Awareness*. No significant main effects or interactions were found for the questions *NotMine*, *TwoHands*, and *Agency*. The data of Experiment 2 met the assumptions for ordinal logistic regressions analyses since the dependent variables are ordinal, the independent variables are categorical, no multi-collinearity was detected (i.e., automatically verified by the Stata software), and Brant tests on each of the repeated measures questionnaire variables showed that the assumption of proportional odds was not violated. Detailed results of the ordered logistic regressions are given in Table 3 and differences between the conditions in the questionnaire items can be seen in Fig. 6.

*Motor performance.* The only significant main effect in the mixed model ANOVA was *Vertical Position of Target*s (F(1,16) = 14.95, p < 0.01, partial $\eta^2$ = 0.48), indicating that independently of the experimental condition, reaction times were higher for touching targets located at the back (M = 561 ms, SD = 90 ms) compared to those located at the front (M = 501 ms, SD = 100 ms) of the virtual table. However, no main effects of *Body Visual Con-*

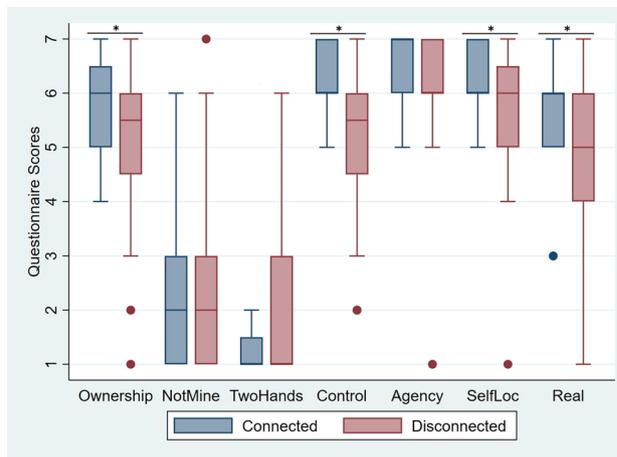

**Figure 6.** Boxplots for the VR questionnaire included in Experiment 2. Boxplots show the median, interquartile ranges, maximum scores, and minimum scores for each questionnaire item in the Connected and Disconnected hands experimental conditions, respectively. Dots represent outlier values and the asterisks show the questionnaire items in which a significant difference between the Connected and Disconnected hands condition was found.

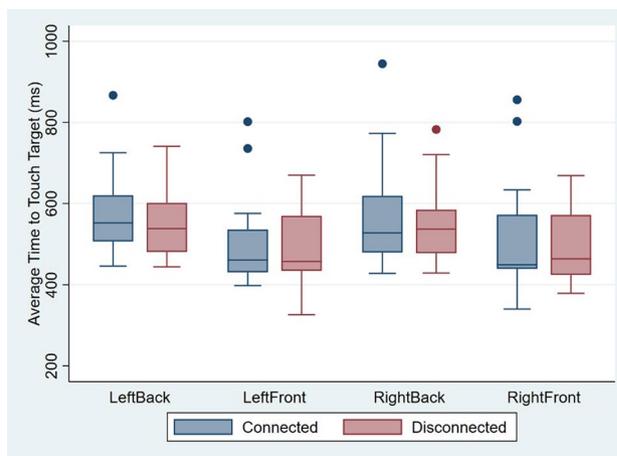

**Figure 7.** Boxplots of the averaged time taken (in milliseconds) to touch a target depending on the target being located on the front, back, right, or left of a virtual table in Experiment 2. The boxplots show the median, interquartile ranges, maximum scores, and minimum scores for the time taken to touch the targets, while the dots represent outlier values.

*tinuity* ($F(1,16) = 1.73$, $p = 0.21$, partial $\eta^2 = 0.10$), *Horizontal Position of Targets* ($F(1,16) = 0.07$, $p = 0.79$, partial $\eta^2 < 0.01$), *Awareness* ($F(1,16) = 0.23$, $p = 0.64$, partial $\eta^2 = 0.14$) or *Order* ($F(1,16) = 0.01$, $p = 0.99$, partial $\eta^2 < 0.01$) were found. Moreover, no significant interaction effects were found between *Body Visual Continuity\*Awareness* ($F(1,16) = 0.01$, $p = 0.99$, partial $\eta^2 < 0.01$) and between *Body Visual Continuity\*Order* ($F(1,16) = 2.54$, $p = 0.13$, partial $\eta^2 = 0.14$). In conclusion, in this second experiment, no evidence was found indicating that motor performance was influenced by the visual connectivity of the virtual hands to a body (Fig. 7).

To discard that differences between Experiment 1 and 2 in motor performance were related to the differences in sample sizes and to further control for the unexpected between subject factor (un)awareness, we ran a post-hoc power analysis for the mixed model ANOVA. Taking into account that in Experiment 1 and 2, the effects sizes were between $\eta^2$ 0.10 and 0.16, we assumed an effect size of $\eta^2 = 0.10$, set alpha to $= 0.05$ and power $= 0.80$. The projected sample size needed with this effect size, based on the GPower software 3.1.9.7, is N $= 20$. We used N $= 31$ participants in the first study and N $= 20$ participants in the second study. Therefore, the sample size suggested by this power analysis is bigger or equal to the included sample sizes in both studies. Therefore, the contradictory findings between Experiment 1 and 2 are unlikely to be related to low power.

*Final interview.* In response to *Question 1*, several participants reported finding the task of touching the virtual cubes fun and engaging. Moreover, other participants said that they were surprised with how accurately the virtual hand reflected their real-time hand movements.

In relation to *Question 2*, in line with the results of Experiment 1, we observed that 10 participants did not explicitly mention the hand connection or disconnection as a difference between the two VR sessions. This means that some participants seemed to not be aware that the forearms of the virtual hands were missing in one of the experimental conditions. 10 participants did explicitly report noticing that in one condition the virtual hands did have forearms and that in the other condition they did not. As in Experiment 1, in order to control for participants being (n = 10) or not being aware (n = 10) of the experimental manipulation, the variable *Awareness* was added as a factor to control for in all the statistical analyses.

**Discussion.** In accordance with the results found in Experiment 1, this experiment indicated that feelings of body ownership, control, and self-location are enhanced with the visual connectivity of the hands to the rest of a body. In this study we further found that participants perceived the virtual hands as less realistic when they were disconnected from the rest of the body. However, it should be noted that the disconnection of the hands did not completely diminish the sense of embodiment, since participants still rated the illusions with relatively high scores. Moreover, in contrast to Experiment 1, our results indicated that being aware or not of the experimental manipulation had an impact on perceived body ownership. Overall, participants who did not notice the experimental manipulation (i.e., connection or disconnection of the hands) reported higher body ownership scores, compared to participants who did notice the different visual appearances of the hands. Finally, in this study where we carefully controlled for task difficulty across conditions and whether participants were aware or not of the experimental manipulation, we were not able to replicate the findings of Experiment 1 in relation to motor performance. In this experiment we did not find significant differences in motor performance when interacting through a connected or disconnected hands.

## General discussion

Several studies have tried to understand the factors that modulate the sense of embodiment of a virtual body. In this regard, a body of evidence has demonstrated that the anthropomorphic features and the degree of visual realism of a virtual body impact the perception of body ownership and agency[4,21,33,38,39]. Under the influence of rich sensory feedback, there is also evidence indicating that the visual appearance of the hands does not play such a prominent role[5,25,40]. For instance, people can experience ownership of a long virtual arm, which is three times longer than their real arm, when they are able to control it through real-time movements[41,42]. Similarly, participants also experience ownership of virtual bodies that are radically different to their real bodies, in aspects such as race[43], transparency[44], age[45], and even gender[46].

A factor that can influence the sense of embodiment is the visual disconnection of the virtual hands or the visual discontinuity of an artificial body. A series of studies has demonstrated that embodiment is diminished when participants observe a virtual hand that is disconnected from the rest of the body[22–24]. However, these studies did not include rich visuomotor feedback that linked the participants' motions to the movements executed by the virtual hands. The above-mentioned studies evoked ownership based on the perception of a collocated virtual static body, thus proprioceptive congruence was mainly exploited to induce embodiment of the artificial limb. Recent evidence has suggested that ownership for disconnected hands can in fact be preserved when rich visuomotor correlations between the real and artificial body are established[25,26]. Nonetheless, these studies only assessed embodiment by means of questionnaire responses and averaged body ownership scores for disconnected limbs were moderately low when compared to past research (e.g. 4.25)[25].

The results of the present study suggest that visuomotor feedback only partially contributes to the preservation of the sense of embodiment for artificial limbs that are detached from the rest of an artificial body. In two independent studies, we have found that subjective feelings of body ownership, control, self-location, and realism are lower for virtual disconnected hands compared to connected hands. This evidence further supports the notion that embodiment arises based on the combination of bottom-up (i.e., congruency of visual, motor, proprioceptive, and tactile inputs) and top-down factors (i.e., visual appearance of artificial body). This is in accordance with results from Tieri et al.[23] and Perez-Marcos et al.[22], which show diminished body ownership for virtual static morphologically incongruent limbs (i.e., disconnected hands). However, the present findings stand in contrast with the results of Brugada-Ramentol et al.[25] and Tran et al.[26]. Despite the need for additional research to understand these contradictory findings, it is possible that differences between these studies and the present experiments are partially explained based on the type of motor task used in VR. In the present studies we included full hand tracking where the motion of the virtual hand and fingers closely matched the real hand movements of the participant, while in Tran et al.[26] finger tracking was not enabled. Moreover, our experiments were based on a bimanual task, while Tran et al.[26] and Brugada-Ramentol et al.[25] used a unimanual motor task. Finally, while tasks from Tran et al.[26] and Brugada-Ramentol et al.[25] were based on the execution of discrete movements during a limited number of trials (e.g., 25 trials), in our experiments we encouraged participants to move their hands freely to touch different virtual cubes during several trials (i.e., more than 100 trials). Therefore, it is possible that the richness of the visuomotor feedback provided (i.e., full hand tracking and using both hands to execute a task), along with the larger number of trials included in our motor tasks, accentuated the perceivable differences between the connected and disconnected hands, leading to a decrease in embodiment feelings for the less realistic ones (i.e., disconnected). In fact, future studies should also research how the richness of different motor tasks impacts the sense of embodiment over an artificial body, since this is a topic that, to our knowledge, has been scarcely investigated[47,48].

It should be noted that despite findings of lower embodiment scores under visual discontinuity conditions, participants still reported quite high scores of body ownership and agency when interacting through the virtual hands detached from a body. This means that it is still possible to experience embodiment of morphologically incongruent limbs, although the sense of embodiment seems to be strengthened by the degree of visual realism and morphological plausibility. This was also supported by preliminary data on physiological responses observed when the virtual hands were threatened, where increased SCR was observed when a threat was presented, independently of the visual appearance of the virtual hands. This stands in contrast with the results of Tieri et al.[24], where skin conductance responses to a threat where modulated by the connectivity of the hands to the rest of the body. However, an important difference between these two studies is the inclusion of real-time visuomotor feedback. In this regard, it is possible that bottom-up information (i.e., visuomotor information) and the perception of agency over the virtual hands lead to an automatic reflex response to avoid a threat, independently of the morphological characteristics of the hands. This is also in accordance with past studies showing strong physiological reactions to a threat when embodying an actively controlled very long arm[41] or different types of non-realistic avatars[49]. However, additional research is required to better understand these effects since the present results are only based on a single trial presentation of the threat per condition and should be interpreted with caution. Future studies should consider including multiple threat-related trials to evaluate whether the same results can be replicated.

Another interesting and unexpected aspect of the present research is that in both experiments we found that half of the participants did not clearly notice the visual connection or disconnection of the virtual hands. However, our results indicate that being aware of the experimental manipulation modulated to some extent body ownership and motor performance. In Experiment 1, we found that participants who did not notice the experimental manipulation achieved a higher number of hits (i.e., touched targets) when interacting through connected hands compared to disconnected. In Experiment 2 we observed that unaware participants had significantly higher body ownership scores compared to participants who noticed the difference in the visual appearances of the hands. Despite the scarcity of the research on this topic, this is not the first study to report the lack of conscious awareness with respect to subjective perception or motor behaviors due to changes observed in an embodied virtual body. In the study of Burin et al.[50], participants did not consciously notice that their drawing motions were influenced by the movements made by a virtual body experienced from first-person perspective. In this experiment, despite participants being instructed to always draw straight lines, they tended to draw ellipses when they saw a virtual collocated body drawing ellipses. However, none of the participants explicitly reported noticing that their motor behavior (i.e., drawing) was influenced by the motion of the virtual body, since they reported that they were always drawing straight lines, even in the condition where they saw the virtual body drawing ellipses. A similar behavior was also recently reported by Gonzalez-Franco et al.[51]. Future research is needed in order to better understand this phenomenon. Moreover, it is important that future studies control for the unawareness or awareness of the experimental differences, when different aspects of bodily perception are manipulated.

No conclusive evidence was found in relation to how body-related incongruent morphological information impacts motor performance in a quick bimanual task. Although Experiment 1 indicated that motor performance was differently influenced by the visual appearance of the virtual hand, these results were not replicated in Experiment 2 which controlled for further variables in the motor task such as target position and the order of presentation of the targets. However, it is interesting to observe that in Experiment 1, our results indicated that unaware participants (i.e., not consciously aware of the experimental manipulation) achieved significantly better motor performance when interacting with highly realistic hands (i.e., connected) compared to morphologically implausible ones (i.e., disconnected). These results were not observed in the case where participants noticed the experimental manipulation and in Experiment 2, making these results difficult to interpret. However, it is interesting to evaluate these preliminary findings in the light of previous research which has shown that bodily actions seem to play a prominent role in shaping body ownership[14,52]. For instance, Della Gatta et al.[53] and Fossataro et al.[13] found that the sense of embodiment for an artificial limb is related to a down-regulation of motor cortex excitability for the real disembodied hand. Moreover, other studies have shown that patients suffering from movement disorders also experience changes in their sense of body ownership[54–56], sometimes leading to stronger illusory experiences for an artificial limb and less ownership for the paralyzed real limb[57]. Consequently, it could be expected that when individuals experience less ownership for an artificial hand, their motor performance might improve as a result of enhanced motor cortex activations and control of the real hand. However, in Experiment 1 we observed the opposite pattern in unaware participants, with increased motor performance when the sense of ownership was stronger due to the embodiment of connected realistic hands. Interestingly, in a recent study, Reader and Ehrsson[58] showed that a decrease in own hand ownership does not necessarily impact motor planning and behaviors. Therefore, it is possible that the feeling of body ownership and its relationship with the motor system are further influenced by additional top-down factors such as body self-consciousness (i.e., noticing differences in the body representation) or even by different affordances related to the visual appearances of an artificial body (e.g., having a long arm might make it seem more flexible or having a disconnected hand might result in it being more difficult to localize in space). This is a topic which deserves further research and which should be addressed by future studies.

The primarily goal of the present study was researching the sense of embodiment for connected or disconnected virtual hands using a natural, engaging, and ecologically valid task, where rich visuomotor feedback was provided. Based on this reason, we opted to use a bimanual task that allowed participants to freely move their virtual hands, instead of using a less natural task requiring simple discrete movements. The use of an artificial task, only requiring restricted and discrete movements, might have potentially prevented participants from fully exploring the richness of the visuomotor feedback (i.e., movements of the virtual hands reflecting real hand movements with very high accuracy). However, future research should investigate the impact of body discontinuity on motor performance under rich visuomotor feedback, for instance using a Fitts' law paradigm, where the size of the targets and positions of the hands are controlled. Moreover, it is important to note that using artificial limbs detached from the rest of an artificial body may not have been the most effective control condition to represent a body part with morphological implausible characteristics in reality, since our results indicate that several participants were not consciously aware of the experimental manipulation. Future studies should consider using different types of control conditions, where the appearance of an artificial body is distorted by manipulating different features (i.e., structure or color of the hand). Alternatively, future research could also make participants explicitly aware of the visual appearance differences between different virtual hands representations.

## Conclusion

In the present study we have found that even under the presence of rich visuomotor feedback, mere observation of body discontinuity diminishes to some extent the sense of embodiment of an artificial virtual body. These results are in agreement with past evidence showing the importance of visual continuity for body ownership, and further support the notion that embodiment arises as a result of the mutual influence of top-down and bottom-up factors. However, our preliminary data indicated that there is no significant impact of visual continuity of a body on physiological reactions to a threat. This might point to a potential role of visuomotor feedback in evoking threat-related responses, despite perceiving morphologically implausible information regarding the body. Finally, contradictory evidence was found in relation to how interaction with virtual disconnected hands impacts motor performance. Moreover, our results also indicate that half of the participants did not consciously notice the inclusion or exclusion of the virtual forearm, despite these visual features being clearly visible. These findings highlight the need for future studies investigating how body visual discontinuity impacts motor behavior and physiological responses. Furthermore, this study also stresses the importance of controlling whether participants consciously notice changes in the visual appearance of their embodied virtual bodies when the visual features are modified.

## Data availability

The data of Experiment 1 and 2 is available as Supplementary Information 1. Possible details identifying participants, such as gender and age, have been removed.

## Acknowledgements

This research has received funding from the European Union's Horizon 2020 research and innovation program under Grant agreement #737087 (Levitate). We want to thank Laura Zeußel, Hannes Pillny, and Matthias Popp for their help in implementing the virtual reality setup and some MATLAB scripts. We also want to thank Jan Milosch and Timm Seltmann for their help in running the study.

### Author contributions

### Funding
Open Access funding enabled and organized by Projekt DEAL. This research received funding from the European Union's Horizon 2020 research and innovation program under Grant agreement #737087 (Levitate).

### Competing interests
The authors declare no competing interests.

### Additional information
**Supplementary Information** The online version contains supplementary material available at https://doi.org/10.1038/s41598-020-79255-5.

**Correspondence** and requests for materials should be addressed to S.S.

**Reprints and permissions information** is available at www.nature.com/reprints.

**Publisher's note**  Springer Nature remains neutral with regard to jurisdictional claims in published maps and institutional affiliations.